\newcommand{\be}{\begin{equation}}      % equation number automatic
\newcommand{\ee}{\end{equation}}
\newcommand{\bea}{\begin{eqnarray}}     % equation number automatic
\newcommand{\eea}{\end{eqnarray}}
\newcommand{\beb}{\begin{eqnarray*}}    % no equation number
\newcommand{\eeb}{\end{eqnarray*}}
 \documentstyle[prl,preprint,aps]{revtex}
\begin{document}
\draft

\title{Superhexatics}
\author{Kieran~Mullen,$^{(1)}$
H.~T.~C.~Stoof,$^{(1),(2)}$ Mats~Wallin,$^{(1),(3)}$
and S.~M.~Girvin,$^{(1)}$}
\address{
$^{(1)}$Physics Department, Indiana University, Bloomington, Indiana 47405\\
$^{(2)}$Institute of Theoretical Physics, University of Utrecht\\
Princetonplein 5, P.~O.~Box 80.006, 3508 TA Utrecht, The Netherlands\\
$^{(3)}$Department of Theoretical Physics \\
Royal Institute of Technology
10044 Stockholm, Sweden \\ }

\maketitle

\begin{abstract}
\baselineskip=14pt
We develop a theory for a novel state of ${}^4$He films, that possesses
off diagonal order (as in the superfluid state), as well as hexatic or
bond orientational order.  Within our description,
both the hexatic and superfluid transitions are still of the
Kosterlitz-Thouless type, but the superfluid stiffness
is sensitive to the hexatic transition.  We briefly discuss the possible
relevance of this work to recent experiments on submonolayer helium
films.

\end{abstract}

\narrowtext

In the early 1970's a substantial amount of theoretical
work was done on the topic
of supersolid
%% FOLLOWING LINE CANNOT BE BROKEN BEFORE 80 CHAR
helium.\cite{{chester},{leggett},{matsuda},{fisherliu},{fishernels},{supersola},{supersolb},{otterlo},{frey},{bruder}}
Such a system would possess both
the quantum coherence of the superfluid state,
characterized by off-diagonal long-range order in the density matrix
(or ``ODLRO''\cite{{penrose}}),
and crystalline order, characterized by long-range order in the
density at some finite wavevector, or ``diagonal'' long range order
in the density matrix.  Progress in the field was hampered by a lack of
convincing experimental evidence that such a state existed in nature.
In this Letter we investigate the properties of  two-dimensional
helium films which have off-diagonal order as well as
hexatic order (characterized by algebraic long-range order
in the orientation of directions to nearest-neighbor atoms).  We call
such a system a ``superhexatic,'' to distinguish it from the superfluid
state which possesses no such orientational order.

Superhexatic systems are interesting for a number of reasons.  First,
the hexatic constraints required on the many-body wavefunction are {\it prima
facie} much weaker than the crystalline ones of a supersolid,\cite{{leggett}}
which should make it easier for the
quantum and spatial order to coexist.
Second,  superhexatics can be mapped to a variety of other
systems of physical interest such as the two-dimensional XY magnet on
a random lattice, vortex lines in superconductors,\cite{{frey}}
or a set of two-dimensional Coulomb charges
in quantum gravity.\cite{{qgrav}}
In such models there are two different species of charge
(for example, point masses and Coulomb charges), each with its
own dynamics and each influencing the other.
Third, the helium system not only provides
an additional theoretical tack, but there is some evidence that
superhexatics may have been seen in experiment.  Recent third sound
measurements
of submonolayer films of helium on hydrogen and deuterium substrates
demonstrate {\it two} independent Kosterlitz-Thouless (KT) transitions: the
standard superfluid transition near
1K and a second one near 0.5K.\cite{{mochela},{mochelb}}
As we shall discuss below, the lower transition may be the freezing
of the  superfluid to form a superhexatic as the temperature is lowered.

We shall show below that
couplings between off-diagonal and hexatic order can exist, and that they
need not destroy the superfluid or hexatic transitions.
These couplings do allow for evidence of the hexatic transition to
be seen in the superfluid stiffness.  We do not address the
microscopic interactions that give rise to these couplings in this work.
 Finally, we make a brief comparison of this work to experiment.

A qualitative\cite{{feynman},{kikuchi},{elser},{kleinert1}} and
quantitative\cite{{cepperley}}
theory of the superfluid phase transition can be built up from the
theory of ``ring exchanges''.  In this
approach the helium atoms possess only their kinetic energy
and a short-range repulsive interaction. The
partition function is calculated using standard path-integral techniques,
in which the system is evolved for an imaginary ``time''
$\hbar\beta\equiv\hbar/kT$, where $T$ is the temperature.
The indistinguishability of the bosons allows for  contributions
to the partition function in which helium atoms have exchanged
positions.  The superfluid transition
occurs when it becomes entropically favorable for
large ``rings'' of bosons to permute
their positions,
establishing phase coherence across the system.\cite{{elser}}

Hexatic order can affect this process in at least two ways.
First, the additional stiffness of the system might reduce
the amplitude for helium atoms to exchange
position,\cite{{leggett},{thouless}}   thereby drastically
lowering the superfluid density.
Second, there is a subtle
topological effect.   The hexatic-fluid transition can be viewed
as a disclination-unbinding transition,\cite{{halpnels},{halprev},{nelsrev}}
where a perfect disclination is a
topological defect in the bond-orientational order.
These point disclinations can interact with the point charges in the
superfluid order.
%Such a defect is characterized by a Burger's vector (the difference in
%length of closed circuits in distorted and  pristine crystal)
%that is proportional in size to the loop of the Burger's circuit.  Since
%the calculated superfluid stiffness depends upon the statistics of
%ring exchanges, and the ring exchanges depend upon the
%of drawing loops on the lattice, the
%superfluid stiffness should be altered by the
%presence or absence of disclinations which change the nature of these loops.
The vortex-disclination interaction appears most naturally in the context of an
XY model in a fluctuating geometry.
Consider, by way of example, a lattice of helium containing a single
disclination.
The theory of ring-exchanges of the atoms can be mapped to a Landau-Ginzburg
theory for the superfluid.
The energy depends only upon gradients of the phase, and can be
related to two-dimensional electrostatics\cite{{halprev},{minnhagen}}
wherein vortices play the role of Coulomb charges.
The effect of the disclination can then be viewed as distorting
the plane containing the charges into either
a cone for a negative disclination, or a saddle for a positive
disclination.\cite{{kleinert2}}
Using conformal mapping one can solve the electrostatics problem for a single
charge on a cone (saddle) and find that it is repelled (attracted) to
the disclination independent of the sign of the charge, and that the
energy of interaction depends logarithmically in their separation.
A similar calculation shows that the energy of a vortex and a dislocation
varies inversely with the distance between them.

It is also possible
to demonstrate couplings between the hexatic and superfluid order using
a microscopic many-body approach,\cite{{longpaper}} but here we will
consider only a phenomenological Landau-Ginzburg model.
The free energy of a superhexatic may be written as a sum of the elastic
energy, the superfluid energy, and an interaction energy.
The elastic term is given by\cite{{kleinert2}}
\bea
E_{\rm el} =
\int  d\vec r\,  {1\over 4\tilde\mu(1+\tilde\nu)}
\left|\nabla^2 \chi(\vec r)\right|^2
   i\, \eta(\vec r) \chi(\vec r)
 +  E_b \vec b (\vec r )^2 +E_\Theta \Theta(\vec r )^2
\label{elastic}
\eea
where $\tilde\mu$ and $\tilde\nu$ are elastic constants,
$\chi(\vec r )$ is the ``gauge field'' of
the stress tensor, $\sigma_{ij}$, so that
$\sigma_{ij}(\vec r)
=\epsilon_{ik}\epsilon_{jl} \partial_k\partial_l \chi(\vec r)$
and $\vec b (\vec r )$ and $\Theta(\vec r )$ are the dislocation
(Burger's vector) and
disclination densities with core energies $E_b$ and $E_\Theta$,
respectively.\cite{corenote}
These defect densities can be combined into a single
scalar, the incompatibility,
$\eta(\vec r)\equiv\Theta(\vec r) + \hat z\cdot\vec\nabla  \times \vec b(\vec
r)$
which acts as a source term for the field
$\chi(\vec r )$, where $\hat z$ is a unit vector normal to the plane of the
film.
We assume that we are well above the melting transition of the
solid, and treat the dislocation density as a continuous
field,\cite{{halpnels}} ignoring the discrete nature of the dislocations.
We next write the superfluid
energy in terms of a ficticious electrostatic potential $\phi(\vec r )$,
related to the phase of the superfluid order parameter, $\theta(\vec r )$,
by $\vec\nabla \phi(\vec r )=-\hat z \times \vec\nabla  \theta(\vec r )$.
Using this potential, the superfluid energy may be written in
a fashion similar to the elastic energy:
\bea
E_{\rm sf}=\int d\vec r\,
{1\over 4\pi \rho_s}\left\vert\vec\nabla  \varphi(\vec r)\right\vert^2
+ i\nu(\vec r ) \varphi(\vec r ) + E_\nu \nu(\vec r )^2,
\eea
where $\nu(\vec r )$ is the density of point vortex ``charges''
for the
field $\varphi(\vec r )$.

The leading order interaction between the two fields may be written as:
\bea
E_{\rm int} =  \int  d\vec r \,\,\,\,
 i\gamma_{jk\ell m}\,\partial_j\partial_k \chi(\vec r )
\partial_\ell\varphi(\vec r)\partial_m\varphi(\vec r)
 +  \lambda_{jk\ell m}\, b_j(\vec r )b_k(\vec r )
\partial_l\varphi(\vec r)\partial_m\varphi(\vec r)
+ E_{\Theta\nu} \Theta(\vec r )\nu^2(\vec r ).
\label{fullinter}
\eea
The coupling tensors $\gamma_{ijk\ell}$ and $\lambda_{ijk\ell}$ are
required to be symmetric  in order to preserve rotational invariance.
They may be split into the traceful and traceless contributions.  For the
purposes of this calculation we consider only the trace (diagonal)
contributions,
\bea
E_{\rm int}^{(0)}  =
\int d\vec r\,\,\, i\gamma_0 \nabla^2 \chi(\vec r )\,
\left|\vec\nabla \varphi(\vec r)\right|^2
+ \lambda_0 \vec b (\vec r )^2 \,
\left\vert\vec\nabla \varphi(\vec r)\right\vert^2
 +  E_{\Theta\nu} \Theta(\vec r )\nu^2(\vec r )
\label{traceinter}
\eea
This is the lowest order interaction one can write for the two fields.
Time reversal invariance places a powerful constraint on the theory:
any interaction requires an even number of
powers of $\vec\nabla \varphi$; there is no similar requirement
on $\vec\nabla \chi(\vec r )$.  The first term in eq.(\ref{traceinter})
represents the coupling of the
stress tensor to the superflow.  In the absence of dislocations
it generates the logarithmic interactions between vortices
and disclinations mentioned earlier.  The second term represents the
possible variation of the local superfluid density (stiffness)
with the local dislocation density.
The final term represents the interaction between the cores of
disclinations and vortices, and can be shown to be generated
from the first term in eq.\ref{traceinter}.

The partition function of the superhexatic system may be written as:
\be
Z=\sum_{\Theta(\vec r ), \nu(\vec r )}
\int {\cal D}\varphi[\vec r] \,{\cal D}\chi[\vec r] \,{\cal D}\vec b [\vec r ]
\,e^{ -\beta (E_{\rm el} + E_{\rm sf} + E_{\rm int}^{(0)})}
\ee
We simplify the partition function in two steps.  First,
we integrate out the gaussian dislocation vector field, $\vec b (\vec r )$,
treating $\lambda_0$ as a small parameter in the resulting action.
This introduces a $|\vec\nabla \chi(\vec r )|^2$ term in the action so that
the interaction of bare disclinations is reduced from
$r^2\log r$ in the absence of dislocations,
to a logarithmic one  due to partial
screening by the dislocations.\cite{{halpnels}}  Second, we will work in the
small fugacity limit, so that we may limit the contributions of the sums
over vortex and disclination charge to $0$ and $\pm 1$,
turning the problem into that of two coupled sine-Gordon
systems.\cite{{minnhagen}}  After some
algebra we obtain:
\bea
Z  =  \int &
   {\cal D}\varphi[\vec r]\, {\cal D}\chi[\vec r]\,
\exp -\beta {\displaystyle\int} d^2r\left\{
{\displaystyle 1\over\displaystyle  4\mu(1+\nu)}
\left\vert\nabla^2 \chi(\vec r)\right\vert^2\right.
 +  \left. {\displaystyle 1\over \displaystyle 2E_b}
\left\vert\vec\nabla \chi(\vec r )\right\vert^2  +
{\displaystyle 1 \over \displaystyle 4\pi \rho_s}
\left\vert\vec\nabla \varphi(\vec r )\right\vert^2 \right. \cr
& + i\gamma_0\nabla^2\chi(\vec r)\left|\vec\nabla \varphi(\vec r )\right|^2
 -  {\displaystyle\lambda_0\over \displaystyle E_b}
\left\vert\vec\nabla \chi(\vec r )\right\vert^2
\left\vert\vec\nabla \varphi(\vec r )\right\vert^2 \cr
+& g_1 \cos\beta\phi(\vec r )+g_2\cos\beta\chi(\vec r )
 +  g_3\cos\beta\chi(\vec r )\cos\beta\phi(\vec r )+
i\,g_4 \sin\beta\chi(\vec r )\cos\beta\phi(\vec r )
  \vphantom{\displaystyle\int}\left.\vphantom{\int}\right\}.
\eea
The constants $g_i$ can be simply related to the original core energies:
$g_1$ and $g_2$ are the fugacities of the vortices and disclinations,
$g_3$ represents the added energy required to place a vortex on a
disclination, and $g_4$ reflects whether or not the energy cost depends
upon the sign of the disclination.

Let  $T_\Theta$  and $T_\nu$  be the
hexatic and superfluid transition temperatures, respectively,
for the uncoupled models ($T_\nu > T_\Theta$).
Elementary power counting indicates that all of the
couplings between the two order parameters allowed by time-reversal symmetry
are irrelevant (in the renormalization-group sense) at the
gaussian fixed point,  $T_\Theta$.  A more rigorous
calculation using momentum shell cutoff renormalization\cite{{knops}}
upholds this conclusion near the line of gaussian fixed points,
even in the regime $T_\Theta< T < T_\nu$, where the disclination
fugacity $g_2$ is perturbatively relevant.\cite{{longpaper}}
This result is quite important:
it shows that weak coupling of the two models will not destroy
the KT nature of either the vortex-unbinding or
disclination-unbinding transition.  It also
allows us to conclude that there will not be a discontinuous jump in the
superfluid stiffness when the hexatic transition occurs.
%At $T_\Theta$ the fugacity and hexatic stiffness renormalize to
%zero, but only as we follow the flow to infinite length scales.
%For small variations above or below $T_\Theta$ we must still carry
%the renormalization analysis
%to large length scales before we can ascertain whether or not the stiffness
%is flowing to zero.  Since the couplings between the models
%flow more rapidly than the stiffness, for any physically reasonable
%value of these coupling, the interactions will have scaled to zero
%before the stiffness has scaled to its final value.  Thus, we do not
%expect to see a jump in the superfluid stiffness at $T_\Theta$.

However, this does not prove that the superfluid stiffness is
wholly insensitive to the presence of the hexatic order.
For example, when we
pass through the hexatic transition we expect a change in
$\left\langle|\vec\nabla \chi |^2\right\rangle$ proportional
to the integral of the bump in the
specific heat.  Such a bump is a non-universal feature of the
transition in that its width and magnitude depend upon detailed
features of the system.\cite{{nelscv}}  If we have a coupling
of the form $|\vec\nabla \chi|^2|\vec\nabla \varphi|^2$,
then we would expect
a change in the superfluid stiffness proportional to this integrated
bump.  For a suitable choice of parameters this can be fairly
sharp.  However, the lowest-order perturbative
renormalization group analysis discussed
above is only valid near the transition, and cannot demonstrate
such a feature.

As a proof-of-principle, we have simulated the superhexatic transition
numerically using a Monte Carlo analysis.  We can define a
hexatic order parameter
$\psi_6(\vec r )=\sum_{\rm neighbors}  e^{i 6\theta_i}
\equiv |\psi_6(\vec r )|\, e^{i\theta_6(\vec r )},$
so that $\theta_6(\vec r )$ plays a role similar to that of the superfluid
phase $\theta(\vec r )$,  and like the phase, its gradient is at right
angles to the gradient of its corresponding
gauge field, $\vec\nabla \chi(\vec r )$,
introduced above.
We simulate the problem as two coupled XY models
on a discrete lattice:
\bea
E = -\sum_{(i,j)} k_1 \cos\big(\theta(r_i)-\theta(r_j)\big)
 +  k_2\cos\big(\theta_6(r_i)-\theta_6(r_j)\big)  \cr
+ s  \left[\cos(\theta(r_i)-\theta(r_j))-1\right]
 \left[\cos(\theta_6(r_i)-\theta_6(r_j))-1\right]
\label{montecarlo}
\eea
For a suitable choice of parameters we can obtain a rapid change
in the superfluid
stiffness as a function of temperature near $T_\Theta$, as shown in fig.(1).

We conclude with a brief comparison of this work with recent experiments
by Cheng and Mochel\cite{{mochela},{mochelb}}
as well as other theoretical approaches to the
same experiments.\cite{{zhang},{kapitul}}
Measurements of third sound velocities in submonolayer ${}^4$He films
on hydrogen and deuterium substrates suggest the existence of two
KT  transitions.  The upper transition occurs near 1K, and is consistent
with the standard superfluid transition on conventional substrates.
The lower one occurs at roughly 0.5K, and is signalled by a
change in the third sound velocity as the temperature is lowered:
for deuterium substrates the velocity drops, whereas for hydrogen
the velocity increases.   The temperature of this transition
scales linearly with the  ${}^4$He density.  For films thicker
than a single monolayer deposited on hydrogen substrates,
an additional third sound mode appears over the temperature range
between the two transitions.

If the lower transition is indeed KT, then it seems likely to be a melting
transition of either a solid or a hexatic.
A superhexatic would have no shear modulus, so that superflow
would not be reduced by pinning, which is consistent with experiment.
Depending upon the details of the coupling between the hexatic
and the superfluid order parameters, it is possible to
obtain an increase or decrease in the superfluid stiffness.  This
change may be sharp, but is not discontinuous, which is also consistent
with the experiments.

The superhexatic model cannot explain the additional third
sound  mode seen in some cases.  However, this mode is only
seen for coverages greater than one monolayer, and may result
from different dynamics in the two layers.  A more crucial
difficulty is dealing with the effect of the substrate potential
on the hexatic transition.  In classical systems a hexagonal
substrate will lock in the hexatic order, so that no transition
occurs.\cite{{halpnels}}  The large zero point motion of both
the helium and the
hydrogen may mitigate this effect, but that must be demonstrated
by further study.

Instead of considering spatial ordering of the helium atoms,
Zhang\cite{{zhang}} and Kapitulnik\cite{{kapitul}} have
independently postulated an ordering of the thermally excited
vortices. In such a picture the
intermediate state would be a
vortex-antivortex lattice ({$\rm V\bar VL$})  with
spontaneously broken time-reversal symmetry.  In this picture the
upper transition is the melting of the lattice to produce a normal
fluid, and the lower transition is the sublimation of the lattice
as its constituents disappear.  Zhang has shown how
the additional third sound branch can be nicely explained in terms
of the optical modes of the {$\rm V\bar VL$}.
However, in such a picture {\it neither}
transition should produce KT behavior for the superfluid.
The upper melting transition would produce a universal jump in the
shear modulus of the {$\rm V\bar VL$}, not in the superfluid density, as is
observed experimentally, and as predicted in our
model. In addition, there is no reason why the additional
branch should only be observed on hydrogen subtrates when the
coverage exceeds one monolayer.

The two pictures differ in that the superhexatic phase should
occur below the new, lower transition, while the {$\rm V\bar VL$}\quad
is predicted to occur at and
above it.  By measuring the vortex diffusivity and the
onset of nonlinear dissipation as a function of temperature,
one should be able to establish which regime exhibits anomalous
ordering.

The above analysis of the interaction of spatial and superfluid
order can be extended to physical systems with
a similar mathematical description.  In particular, it can be applied
to an array of Josephson junctions that has been deliberately
constructed to include a disclination or dislocation.  The geometric
analysis discussed above would predict a force on vortices that
attracts or repels them from the defect center. Such an interaction
might be visible in experiments on the ballistic motion of vortices.

In conclusion, we have introduced the notion of a novel phase for ${}^4$He
films called a ``superhexatic'',
which displays both off-diagonal and bond-orientational
order.  We have shown that couplings between the two
exist, and that the couplings preserve the KT nature of the
two transitions.  We have shown that this phase is roughly consistent
with experiments, and suggested ways to test if this approach is
correct.  Study of the system dynamics is underway.\cite{{longpaper}}

\acknowledgments

This research was  supported
by grant NSF DMR-9113911 from the National Science Foundation.
We are grateful to  D.~Arovas, K.~Burke, D.~Huse, J.~Mochel, S.~Renn,
S.~Sondhi, A.~Stern and S.-C.~Zhang for helpful discussions.
We thank A.~Stern for the suggestion
that topological effects could be studied in Josephson junction arrays.

\begin{figure}
\caption[]{ The jump in the superfluid   stiffness during the hexatic
transition, as determined by a Monte Carlo simulation of the action
in eq.(\ref{montecarlo}) for a 100x100 lattice with periodic
boundary conditions.  The size and
width of the jump are proportional to that
of the integral of the specific heat of the hexatic over the transition.
The parameters $K_1=1$ and $K_2=2$,
were chosen so that the uncoupled hexatic and superfluid
transitions occur at 1 and 2, in these dimensionless units.  The
stiffness is calculated for coupling $S=0$ (circles) and
$S=0.5$ (squares).  \label{jumppic}}
\end{figure}

\end{document}